\documentclass[12pt]{article}
\usepackage[english]{babel}
\usepackage{times}
\usepackage[T1]{fontenc}
\usepackage{epsfig}
\usepackage{amsmath}
\usepackage{amssymb}
 
\begin{document}
 \numberwithin{equation}{section}
\title{Relativistic Quantum Mechanics \\ of the Majorana Particle
\thanks{Article based on  lecture at LIX Cracow School of Theoretical Physics,   Zakopane, 2019.}
}
\author{H. Arod\'z \\
{\small Jagiellonian University, Cracow, Poland} \footnote{henryk.arodz@uj.edu.pl}
}

\date{$  $}
\maketitle

\begin{abstract}
This article is a pedagogical  introduction to relativistic quantum mechanics of the free Majorana particle. This  relatively simple theory differs from the well-known quantum mechanics of the Dirac particle in several important aspects.   We present its three equivalent formulations. Next,  so called axial momentum  observable is introduced, and  general solution of the Dirac equation is discussed in terms of eigenfunctions of that operator. Pertinent irreducible representations of the Poincar\'e group are discussed.  Finally, we show  that in the case of massless Majorana particle the quantum mechanics can be reformulated as a spinorial gauge theory. 
\end{abstract}
\pagebreak
  
\section{Introduction}
The concept of Majorana particles is very popular in particle physics  nowadays.  The physical object of interest is a spin 1/2,  electrically neutral, fermionic particle, which does not have its anti-particle. The common spin 1/2 particles like electrons or quarks do possess anti-particles. The Majorana particles are hypothetical objects as yet,  but it is not excluded that there exist neutrinos of this kind \cite{giunti}, \cite{bilenky}.  Anyway,  theory  of such particles is interesting on its own right.  The Majorana (quasi-)particles are intensely studied also  in condensed matter physics, but we shall not touch upon this line of research.

The theory of Majorana particles can be developed on two levels:  as quantum field theory or relativistic quantum mechanics. 
Of course the state of art approach is the field theoretic one. Nevertheless,  the relativistic quantum mechanics also offers some advantages, in particular it is much simpler -- field theory is hard to use except  some rather narrow range of problems like scattering  processes at energies lying within a perturbative regime. The  usefulness of the relativistic quantum mechanics is well documented in the  theory of electrons,  either bound in atoms or traveling in space.  The electron is  the example of Dirac particle.  Relativistic quantum mechanical theories of the Majorana  and the Dirac particle are significantly different. The Dirac particle is a well understood textbook item,  as opposed to the Majorana particle where several subtleties are present. 

In this paper, we attempt to give a pedagogical introduction to  relativistic quantum mechanics of the free Majorana particle. 
It is not comprehensive,  we rather focus on selected topics: the problem of momentum observable;  the general solution  of the Dirac equation  for the Majorana bispinor;  and relativistic invariance in terms of  representations of the Poincar\'e group.  We also describe in detail the  path  from quantum mechanics of the Dirac particle  to  quantum mechanics of the Majorana particle. We emphasize the fact that  in the case of Majorana particle  the quantum mechanics  employs only the algebraic field of real numbers $\mathbb{R}$, while in the Dirac case the complex numbers
are essential. Such real quantum mechanics is less known, but it is thoroughly discussed in literature, see, e.g.,  \cite{varad}, \cite{wootters},   \cite{myr}.   There is an interesting aspect of the theory of massless Majorana particle, namely a local gauge invariance  in the momentum  representation for bispinors, presented in Section 5.2.   To the best of our knowledge, such gauge invariance has not been discussed in literature.

Few words about our conventions.  We use the natural units  $c=\hbar=1$.  Metric tensor $(\eta_{\mu\nu})$ in the Minkowski space-time is  diagonal with the entries $(1, -1,-1,-1)$. Summation over repeated indices is understood.  Four-vectors  and  three-component vectors  have  components with upper indices, for example $ p=(p^0, p^1, p^2, p^3)^T$ or  $\mathbf{x} = (x^1, x^2, x^3)^T$,  unless stated otherwise.  
 Three-component vectors are denoted by the boldface.   $T$ denotes the matrix transposition. In the matrix notation, $\mathbf{x}$ is a column with three, and $p$ with four elements. Bispinors are columns with four elements. 
For convenience, we do not avoid complex numbers when it is natural to use them. For example, we stick to the standard notation for the Dirac matrices $\gamma^{\mu}$.  In the Majorana quantum mechanics they are imaginary, hence we use the real matrices $ i\gamma^{\mu}$, where $i$ is the imaginary unit. Of course, we could get rid of the complex numbers completely at the price of introducing a new notation. 

The paper is organized as follows. In Section 2 we introduce charge conjugation and we define the Majorana bispinors. Section 3 is devoted to the momentum observable for the Majorana particle. In Section 4 we study  general solution of the Dirac equation in the case of Majorana particle.   Relativistic invariance and pertinent representations of the Poincar\'e group are discussed in Section 5.   Remarks are collected in Section 6. 

\section{ The Majorana bispinors and the  Majorana  mass term }
Let us begin from  the Dirac equation for  complex four-component  bispinor $\psi(x)$
 \begin{equation}
 i \gamma^{\mu}_D ( \partial_{\mu} + i q A_{\mu}(x)) \psi(x) - m \psi(x) =0,
 \end{equation}
where  $m $ and $q$ are real constants, the index $\mu$ takes  values 0, 1, 2 and 3,    $A_{\mu}(x)$ is a fixed   four-potential of the  electromagnetic field. The argument of $\psi$,  $x =(t, \mathbf{x})$, denotes points in the Minkowski space-time.  

The  matrices $ \gamma^{\mu}_D$  have the following form 
\begin{equation}
 \gamma^{0}_D=\left( \begin{array}{cc} \sigma_0 & 0 \\ 0 &- \sigma_0 \end{array}      \right), \;\;\;\;     \gamma^{i}_D = \left(\begin{array}{cc} 0 & \sigma_i \\ -\sigma_i & 0 \end{array}     \right),
\end{equation}
known as the Dirac representation. Here $\sigma_i$ are the Pauli matrices, the index $i$ takes  values 1, 2 and 3,  $\sigma_0$  denotes the two-by-two unit matrix.  The matrices $\gamma^{\mu}_D$ obey the condition
\begin{equation}
\gamma^{\mu}_D \gamma^{\nu}_D  + \gamma^{\nu}_D \gamma^{\mu}_D = 2 \eta^{\mu\nu} I,
\end{equation}
where $I$ denotes four-by-four unit matrix.  Note that $\gamma^2_D$ is imaginary while the remaining matrices are real. 

The information given above is purely mathematical. The physical  meaning  of it is established by interpreting $\psi(x)$ as the wave function of certain particle.  The constants $m$ and $q$ then give, respectively,   the rest mass and the electric charge of this particle  (in fact the rest mass is given by $|m|$, not by $m$).  Equation (2.1) and such interpretation of $\psi$  are the basic ingredients of the theory called the quantum mechanics of the Dirac particle. It is  the most important example of relativistic quantum mechanics. Scalar product of two wave functions $\psi_1, \psi_2$  -- necessary in quantum mechanics -- has the form 
\begin{equation}
 \langle \psi_1 | \psi_2 \rangle =  \int\!d^3x \:\overline{\psi}_1(t, \mathbf{x}) \gamma^0_D \psi_2(t, \mathbf{x})  =
 \int\!d^3x \:\psi^{\dagger}_1(t, \mathbf{x}) \psi_2(t, \mathbf{x}), 
 \end{equation}
where  $\overline{\psi} = \psi^{\dagger}\gamma^0_D$ and $\dagger$ denotes the Hermitian conjugation.  The bispinors $\psi$
are  columns  with four elements,  and $\overline{\psi}$'s are one-row matrices with four elements. One can prove that the scalar product  (2.4) does not depend on time $t$  provided that $\psi_1, \psi_2$ are solutions of the Dirac equation (2.1).  

Alternative interpretation of $\psi(x)$, which is not used here,  is that it is a classical field   known as the Dirac field. 

Charge conjugate bispinor $\psi_c(x)$ is defined as follows
\begin{equation}
\psi_c(x) =  i \gamma^2_D \psi^*(x),
\end{equation}
where $^*$ denotes the complex conjugation.  Taking  complex conjugation of Eq.\ (2.1) and  using the relation  $\gamma^2_D (\gamma^{\mu}_D)^* \gamma^2_D = \gamma^{\mu}_D$  we obtain the equation 
\begin{equation} 
 i \gamma^{\mu}_D ( \partial_{\mu} - i q A_{\mu}(x)) \psi_c(x) - m \psi_c(x) =0,
 \end{equation}
 which differs from  Eq.\  (2.1) by the sign in the first term. In consequence,  $\psi_c(x)$ is the wave function of another Dirac particle which has  the electric charge $-q$.

Let us now consider  the Poincar\'e  transformations of  the  Cartesian coordinates in the Minkowski space-time,  $x'= L x +a$. The corresponding transformations of the   bispinor $\psi(x)$ have the form  \begin{equation} \psi'(x) = S(L) \psi(L^{-1}(x-a)), \end{equation}
where  $S(L) = \exp(\omega_{\mu\nu} [\gamma^{\mu}_D, \gamma^{\nu}_D]/8)$. The bracket $[\;,\; ]$ denotes the commutator of the matrices.  The real numbers $\omega_{\mu\nu}= - \omega_{\nu\mu}$ parameterize the proper orthochronous Lorentz group in a vicinity of the unit element $I$, namely $L=\exp(\omega^{\mu\;\;}_{\;\;\nu})$, where $\omega^{\mu\;\;}_{\;\;\nu} = \eta^{\mu\lambda} \omega_{\lambda\nu}$.  
Using  definition (2.5) and formula  (2.7) we find that $\psi_c$ has the same transformation law as $\psi$:
 \begin{equation} \psi'_c(x) = S(L) \psi_c(L^{-1}(x-a)). \end{equation}
This fact inspired E. Majorana  \cite{maj} to proposing an interesting modification of the  quantum mechanics of the Dirac particle. 

The modification consists in generalizing equation (2.1) by including the term  $m_M \psi_c(x)$, often called the Majorana  mass term, 
 \[
  i \gamma^{\mu}_D ( \partial_{\mu} + i q A_{\mu}(x)) \psi(x) - m \psi(x)  - m_M \psi_c(x)  =0, 
 \]
 where   we assume for simplicity that the constant $m_M$ is a real. 
 Such modification however can not be done without a price.  We know from classical electrodynamics  that all four-potentials   which differ by a gauge transformation are physically equivalent, that is $A_{\mu}(x)$  is equivalent  to  $A’_{\mu}(x)= A_{\mu}(x) + \partial_{\mu}\chi(x),$  where $\chi(x)$ is arbitrary smooth real function which vanishes quickly when $x \rightarrow \infty$ (in mathematical terms, it is a test function of the Schwartz class).  
So let us write Eq.\  (2.1) with $A'_{\mu}(x)$
 \begin{equation}
 i \gamma^{\mu}_D ( \partial_{\mu} + i q A'_{\mu}(x)) \psi'(x) - m \psi'(x) =0,
 \end{equation}
where $\psi'(x)$ denotes solutions of this new equation. It is clear that this equation is equivalent to (2.1) -- it suffices to substitute $ \psi'(x) = \exp(-iq \chi (x)) \psi(x)$ and to divide both sides of  equation (2.9)  by  $\exp(-i q \chi (x))$. We say that 
Eq.\  (2.1) is gauge invariant. 
 The gauge invariance is lost when we include  the Majorana  mass term. The reason is that  $ \psi'_c(x) = \exp(i q \chi (x)) \psi_c(x)$, as  follows from definition (2.5), and  therefore the exponential factors can not be removed.  The Majorana mass term breaks the gauge invariance. Therefore such mass term can be considered only if $q=0$, that is when the particle is electrically neutral. 
 For such electrically  neutral particle we may consider the equation
 \begin{equation}
   i \gamma^{\mu}_D  \partial_{\mu}  \psi(x) - m \psi(x)  - m_M \psi_c(x)  =0, 
\end{equation}
known as the Dirac equation with the Majorana mass term.  

The inclusion of the Majorana mass term has a deep implication for the structure of the theory -- it partially breaks the superposition principle of quantum mechanics of the Dirac particle.  The original Dirac equation 
(2.1) is linear over $\mathbb{C}$, that is any  complex linear combination  of its solutions  also is a solution. 
Because $\psi_c$ involves the complex conjugation,  Eq.\ (2.10) allows for linear combinations with real coefficients only. 
On the other hand,  the bispinor $\psi$ is still complex, that is the Hilbert space of the wave functions is linear over $\mathbb{C}$. 
It is clear that one can avoid this discrepancy by taking a smaller Hilbert space in which only real linear combinations are allowed.  The crucial condition for such restriction is that it should be compatible with the Poincar\'e invariance.

Equation (2.10) can be transformed into  equivalent equation for $\psi_c$, namely
\begin{equation}
   i \gamma^{\mu}_D  \partial_{\mu}  \psi_c(x) - m \psi_c(x)  - m_M \psi(x)  =0.  
\end{equation}
Let us introduce new bispinors  $\psi_{\pm}(x)= \frac{1}{2}(\psi(x) \pm \psi_c(x))$.  It follows from Eqs.\ (2.10), (2.11) that
\begin{equation}
   i \gamma^{\mu}_D  \partial_{\mu}  \psi_{\pm}(x) - ( m  \pm m_M) \psi_{\pm}(x)  =0.  
\end{equation}
These equations for $\psi_{\pm}$  have the Dirac  form  (2.1)  (with $q=0$),  but the rest masses are different  if $m \neq 0$ and $m_M \neq 0$,  namely $m_+ = |m + m_M|, \; m_- = |m - m_M|$.   Thus, instead of single equation (2.10) we now have two independent equations  (2.12).  The bispinor $\psi$ is split into the $\psi_{\pm}$ components:  $\psi(x) = \psi_+(x) + \psi_-(x)$.  It turns out that also  scalar product  (2.4)  is split, 
\[ \langle \psi_1 | \psi_2 \rangle = \langle \psi_{1+}| \psi_{2+} \rangle  +     \langle \psi_{1-}| \psi_{2-} \rangle. \]
The Poincar\'e  transformations of  $\psi_{\pm}$ have the same form as for $\psi$ or $\psi_c$, cf. formulas  (2.7), (2.8). 
To summarize,  quantum mechanics of the Dirac particle with the Majorana mass term has been split into two independent  sectors. The splitting is preserved by  the Poincar\'e  transformations. 

The components  $\psi_{\pm}$ are  characterized by their behavior under the charge conjugation. The component $\psi_+$ is charge conjugation even while $\psi_-$ is  odd, namely
\begin{equation} (\psi_+)_c(x) = \psi_+(x), \;\;\;  (\psi_-)_c(x) = - \psi_-(x). \end{equation}
The decomposition of $\psi$ into even and odd components is unique: if  $\psi= \chi_+ + \chi_-$, where $\chi_+$ is even and $\chi_-$ odd, then one can easily show that $\chi_+= \psi_+, \; \chi_- = \psi_-$. 
Conditions (2.13)  define two subspaces of bispinors which are  linear spaces over $\mathbb{R}$, not over $\mathbb{C}$. 
For example, let us consider a linear combination of two charge conjugation even bispinors  $c_1 \psi_+ + c_2 \chi_+$.  The charge conjugation acting on it gives  
\[ (c_1 \psi_+ + c_2 \chi_+)_c = c_1^* \psi_+ +  c_2^* \chi_+. \] 
Thus, the linear combination is charge conjugation even only if $c_1, c_2$ are real numbers. 

The relativistic quantum mechanics of the  Majorana particle is obtained by  taking only the charge conjugation even sector. 
In accordance with considerations presented above,  the Hilbert space consists of (in general) complex bispinors  --  we now denote them by $\psi$ instead of $\psi_+$   ---  which  obey the condition  \begin{equation}\psi_c(x) = \psi(x).\end{equation}   This Hilbert space is linear over $\mathbb{R}$. The scalar product still has the form  (2.4). Time evolution of $\psi$ is governed by the Dirac equation (2.12), in which we rename $m + m_M$ to $m$.  Here we consider only the free Majorana particle. 
More general theory can be obtained by including a certain fixed potential  in the Dirac equation. 

Condition (2.14) can be solved. To this end, we write $ \psi = \left(\begin{array}{c} \xi \\ \zeta  \end{array} \right), $
where $\xi, \zeta$ are two-component spinors. Using definition (2.5) and the explicit form of $\gamma^2_D$  given by (2.2) we find  that  $\zeta = - i \sigma_2 \xi^*$. Therefore, 
\begin{equation}
\psi = \left(\begin{array}{c} \xi \\ - i \sigma_2 \xi^* \end{array} \right),
\end{equation}
where $\xi$ is arbitrary complex spinor. The scalar product of  $\psi$ and $\chi =  \left( \eta,  - i \sigma_2 \eta^*   \right)^T$
is expressed by $\xi$ and $\eta$, 
\begin{equation} 
\langle \psi| \chi \rangle = \int \!d^3x \: \left( \xi^{\dagger} \eta + \eta^{\dagger} \xi\right). 
\end{equation}
The Dirac equation is equivalent to the following equation for the spinor $\xi$
\begin{equation}  i \partial_0 \xi(x) + \sigma_i \sigma_2 \partial_i \xi^*(x) - m \xi(x) =0.   \end{equation}
Formulas (2.15), (2.16) and Eq.\ (2.17)  constitute the so called two-component  formulation of the quantum mechanics of the Majorana particle.  It is used, for example, in \cite{wiese}.

Yet another formulation is obtained  by decomposing the spinor $\xi$ into real and imaginary parts, $\xi = ( \xi' + i \xi'')/\sqrt{2}$, and
rewriting formula (2.15)  in the following form
\begin{equation}
\psi(x) = \frac{1}{\sqrt{2}} \left( \begin{array}{cc}  \sigma_0 & i \sigma_0 \\ - i \sigma_2 & - \sigma_2 \end{array} \right)\: \left( \begin{array}{c}  \xi'  \\ \xi'' \end{array} \right). \end{equation}
The coefficient $1/\sqrt{2}$ is introduced for  convenience. 
The four-by-four matrix on the r.h.s. of   formula (2.18) is non singular -- in fact it is unitary.  Therefore, the Dirac equation for $\psi$ can be equivalently rewritten as  equation for the real bispinor $\Xi = (\xi', \xi'')^T$.  This new equation also has the form of  Dirac equation
\begin{equation}
  i \gamma^{\mu}_M  \partial_{\mu}  \Xi(x) -  m  \Xi(x)  =0, 
  \end{equation}
 with the following  matrices  $\gamma^{\mu}_M$ in place of  $\gamma^{\mu}_D$:
 \begin{equation*}
 \gamma^0_M = i \left( \begin{array}{cc}  0 &  \sigma_0 \\ - \sigma_0 & 0 \end{array} \right), \;\; \gamma^1_M = - i \left( \begin{array}{cc}  0 &  \sigma_3 \\  \sigma_3 & 0 \end{array} \right),
 \end{equation*}
\begin{equation}  \end{equation}
\[   \gamma^2_M = i
  \left( \begin{array}{cc} - \sigma_0 &  0 \\ 0 &  \sigma_0 \end{array} \right), \;\;  \gamma^3_M = i \left( \begin{array}{cc}  0 &  \sigma_1 \\  \sigma_1 & 0 \end{array} \right). \]
  These matrices are unitarily equivalent to the matrices $\gamma^{\mu}_D$. 
Note that all  matrices $\gamma^{\mu}_D$ are purely imaginary \footnote{In such a case we say that we have  the Majorana representation for  $\gamma^{\mu}$ matrices.}. They of course satisfy the Dirac condition (2.3).   For the scalar product  we obtain 
\begin{equation}
\langle \psi_1| \psi_2 \rangle =  \int\! d^3x \:\left(\xi_1^{'T}(x) \xi_2'(x) + \xi_1^{''T}(x) \xi_2''(x)\right) =   \int\! d^3x \: \Xi_1^T(x)  \Xi_2(x),
\end{equation}
where $\Xi_1$  $( \Xi_2)$ corresponds to $\psi_1$ $(\psi_2)$.  In the remaining part of this article we will use  this last formulation.  

Quantum mechanics with  (bi)spinorial wave functions  is  also  used in theory of the Weyl particle.    Relations between the  Dirac, Majorana, and Weyl  quantum particles are elucidated in, e.g., \cite{pal}.

\section{The axial momentum}
Let us dig a bit deeper into the relativistic quantum mechanics of the Majorana particle.  We  will use the third formulation presented above.
In order to facilitate the considerations we now adjust the notation and list the basic tenets of the theory.  From now on, the Majorana real bispinor is denoted by $\psi$ instead of $\Xi$. As the Dirac matrices  in the Majorana representation  we take (for a personal reason) the following matrices
 \[ \gamma^0 = \left(  \begin{array}{cc} 0 & \sigma_2 \\ \sigma_2 & 0 \end{array}  \right), \;\;    \gamma^1 = i \left(  \begin{array}{cc} - \sigma_0 & 0 \\ 0 & \sigma_0 \end{array}  \right), \;\;   \gamma^2 = i \left(  \begin{array}{cc} 0 & \sigma_1 \\ \sigma_1 & 0 \end{array}  \right),\] \begin{equation} \end{equation} \[  \gamma^3 = -i \left(  \begin{array}{cc} 0 & \sigma_3 \\ \sigma_3 & 0 \end{array}  \right),  \;\;\;\;\;   \gamma_5 = i \gamma^0 \gamma^1 \gamma^2 \gamma^3 =  i \left(  \begin{array}{cc} 0 & \sigma_0 \\ - \sigma_0 & 0 \end{array}  \right),  \]
which are unitarily equivalent to the matrices $\gamma^{\mu}_M$ . The matrices $\gamma^0, \gamma_5$ are Hermitian and anti-symmetric,  $\gamma^i$ are anti-Hermitian and symmetric.  The  pertinent Hilbert space ${\cal H}$ consists of all normalizable  real bispinors. It is linear space over $\mathbb{R}$, not over $\mathbb{C}$.   The scalar product is defined as follows
\begin{equation} \langle \psi_1| \psi_2 \rangle =  \int\! d^3x \: \psi_1^T(t, \mathbf{x})  \psi_2(t, \mathbf{x}). \end{equation}  
 Observables are represented by linear operators which are Hermitian with respect to this scalar product. 
Time evolution of the real bispinors is governed by the Dirac equation
\begin{equation}
   i \gamma^{\mu}  \partial_{\mu}  \psi(x) - m \psi(x)   =0, 
\end{equation}
with  imaginary $\gamma^{\mu}$ matrices (3.1).  It is convenient to rewrite this equation in the Hamiltonian  form
 \begin{equation}
 \partial_t \psi = \hat{h} \psi, 
 \end{equation}
where 
\[ \hat{h} = - \gamma^0 \gamma^k \partial_k - i m \gamma^0. \]
This operator is real, but it is not Hermitian.  Nevertheless, the scalar product  turns out to be constant in time because $\hat{h}$ is anti-symmetric as operator  in ${\cal H}$, that is  \[  \langle \psi_1| \hat{h} \psi_2 \rangle =  -  \langle \hat{h} \psi_1| \psi_2 \rangle . \] 
We shall study solutions of Eq.\ (3.4) in the next Section. 

The quantum mechanical  framework described above has certain unusual features. First, the Hamiltonian $\hat{h}$ is not Hermitian, hence it is not an observable. Let us stress that it is not a disaster for the quantum mechanics -- what really matters is 
constant in time scalar product.  Simple calculation shows that scalar product (3.2) is constant in time provided that $\psi_1, \psi_2$ obey equation ( 3.4).  Of course,  the question arises whether there is certain Hermitian energy operator.   The form of general solution  of Eq.\ (3.4) presented in the next Section, see formula (4.5), suggests  the operator \[ \hat{E} = \sqrt{m^2 - \nabla^2}. \]

In the present Section  we   focus on another peculiarity:  the standard momentum operator  $\hat{\mathbf{p}} = - i \nabla$
turns real bispinors into imaginary ones, hence it is not  operator in  the Hilbert space ${\cal H}$. To the best of our knowledge, this problem was noticed first in \cite{pedro} and later readdressed in \cite{arodz}. 
 Is there a replacement for $\hat{\mathbf{p}}$? 
The momentum operator is usually associated with transformation of the wave function $\psi$ under spatial translations, $ \psi'(\mathbf{x}) = \psi(\mathbf{x} - \mathbf{a})$, where $\mathbf{a}$ is a constant vector.  For infinitesimal translations  \[ \psi'(\mathbf{x}) = \psi(\mathbf{x}) - (\mathbf{a} \nabla)\psi(\mathbf{x})  +  {\cal O}(\mathbf{a}^2). \]
Thus, the actual generator of translations is just the $\nabla$ operator, but it is not Hermitian.  When  complex numbers are allowed we multiply $\nabla$ by $-i$ in order to obtain the Hermitian operator $\hat{\mathbf{p}}$. Then we have
\[ \psi'(\mathbf{x}) = \psi(\mathbf{x}) - i  (\mathbf{a}\hat{\mathbf{p}})\psi(\mathbf{x})  +  {\cal O}(\mathbf{a}^2). \]
Below we give an argument that in the Majorana case the natural choice is to multiply $\nabla$ by the matrix $-i\gamma_5$.  This gives
the Hermitian operator $\hat{\mathbf{p}}_5 = - i \gamma_5 \nabla$, called  by us the axial momentum.  In this case
\[ \psi'(\mathbf{x}) = \psi(\mathbf{x}) - i \gamma_5 (\mathbf{a} \hat{\mathbf{p}}_5)\psi(\mathbf{x})  +  {\cal O}(\mathbf{a}^2), \]
because $\gamma_5^2 = I$. 

The argument  for $\hat{\mathbf{p}}_5$ is as follows.  There exists a mapping  between  the Majorana bispinors $\psi$ and   right-handed (or left-handed)  Weyl bispinors $\phi$,  namely  $\psi =  \phi + \phi^*$.  By the definition of right-handed bispinors, $\gamma_5 \phi = \phi$.    It follows that $\gamma_5 \phi^* = - \phi^*$. Therefore  $\gamma_5\psi = \phi - \phi^*$ and 
   $ \phi = (I+\gamma_5) \psi /2 $,  $\phi^* = (I - \gamma_5)\psi/2$.   We see that the mapping is invertible.     
Now, the momentum operator $\hat{\mathbf{p}} = - i \nabla$ is well-defined for the Weyl bispinors because they are complex.  
Moreover,  because $\hat{\mathbf{p}}$ commutes with $\gamma_5$, also  $\hat{\mathbf{p}}\phi$   is  right-handed Weyl  bispinor. 
  Let us find the Majorana bispinor that corresponds to $\hat{\mathbf{p}}\phi$:   \[  \hat{\mathbf{p}} \phi  + (\hat{\mathbf{p}}\phi)^* = - i \nabla (\phi - \phi^*)  = - i \nabla \gamma_5 ( \phi + \phi^*) =  \hat{\mathbf{p}}_5 \psi.\] Thus, the axial momentum operator  in the space of Majorana bispinors  corresponds to  the standard momentum operator in the  space of right-handed Weyl bispinors. 

Normalized eigenfunctions $\psi_{ \mathbf{p}}(\mathbf{x})$ of  the axial  momentum  obey  the equations 
\[  \hat{\mathbf{p}}_5 \psi_{ \mathbf{p}}(\mathbf{x}) =  \mathbf{p}\:  \psi_{ \mathbf{p}}(\mathbf{x}),  \;\;\;\; \int\!d^3x\:  \psi^{T}_{ \mathbf{p}}(\mathbf{x})\: \psi_{\mathbf{q}}(\mathbf{x}) = \delta(\mathbf{p} - \mathbf{q}),  \] 
and they have the following form 
\begin{equation}
 \psi_{ \mathbf{p}}(\mathbf{x}) =  (2\pi)^{-3/2} \exp(i \gamma_5 \mathbf{p}\mathbf{x}) \: v.  
\end{equation}
Here  $v$ an arbitrary real, constant, normalized ($ v^T v=1$) bispinor. For the exponential we may use the formula  
\[  \exp(i \gamma_5 \mathbf{p}\mathbf{x}) =  \cos(\mathbf{p}\mathbf{x}) I  + i \gamma_5 \sin( \mathbf{p}  \mathbf{x} ). \] 
 The eigenvalues $\mathbf{p}$  take arbitrary real values.

The axial momentum is not constant in time in the Heisenberg picture when $m\neq0$. This is  rather unexpected feature, recall that we consider a free particle. Let us first introduce the Heisenberg picture.  Equation   (3.4) has the formal solution 
\[  | t \rangle = \exp(t \hat{h}) \:|t_0 \rangle,  \]   where  $|t_0\rangle$ is an initial state. Time dependent  expectation value of an observable $\hat{{\cal O}}$  is given by
\[ \langle t |   \hat{{\cal O}} |t \rangle  = \:\langle t_ 0 | \exp(- t \hat{h}) \: \hat{{\cal O}}\:\exp(t \hat{h}) |t_0 \rangle .\]  Therefore, we define  the Heisenberg picture version of  $\hat{{\cal O}}$  as
\[  \hat{{\cal O}}(t) =  \exp(- t \hat{h})\: \hat{{\cal O}}\: \exp(t \hat{h}). \] 
In consequence 
\begin{equation}
\frac{d  \hat{{\cal O}}(t)}{ d t} =  \left[  \hat{{\cal O}}(t),   \hat{h} \right] + (\partial_t  \hat{{\cal O}})(t),
\end{equation}
where the last term on the r.h.s.  appears when  $ \hat{{\cal O}}$ is   time dependent in the Schroedinger picture. 
 In the case of axial momentum  the r.h.s. of  Eq.\ (3.6) does not vanish when $m \neq0$, 
 \[
 [ \hat{\mathbf{p}}_5, \hat{h}]= 2 i m \gamma^0 \hat{\mathbf{p}}_5. 
 \]
Solution of the Heisenberg equation (3.6)  reads \cite{arodz}
 \begin{equation}  \hat{\mathbf{p}}_5(t) = - i \gamma_5(t) \nabla, \end{equation} 
 where
 \begin{equation}  \gamma_5(t) = \gamma_5 + i m \hat{E}^{-1}\gamma^0 \gamma_5 \left[ \sin(2 \hat{E}t)  +    \hat{J}  (1- \cos(2 \hat{E}t))\right],  \end{equation}
and  $\hat{J} = \hat{h}/\hat{E}$.   Because  $\hat{J}^2 = - I$,   
the two oscillating terms  on the r.h.s of formula (3.8)  are of the same order  $m/\hat{E}$.  

Notice that $\hat{\mathbf{p}}_5^2 = - \nabla^2$ commutes with $\hat{h}$. Therefore the energy $\hat{E}$ as well as $|\hat{\mathbf{p}}_5|$ are constant in time.  The evolution of $ \hat{\mathbf{p}}_5(t)$ reminds a precession.

Matrix elements of the axial momentum can depend on time, for example
\[ \int\!\! d^3x \: \psi^T_{\mathbf{p}}(\mathbf{x}) \hat{\mathbf{p}}_5(t)\psi_{\mathbf{q}}(\mathbf{x}) = \mathbf{p} \left[1 + \frac{m^2}{E_p^2}(\cos(2 E_p t)-1) \right] (v^Tw) \:\delta(\mathbf{p}-\mathbf{q})    \]
\[- \mathbf{p}  \frac{ m}{E_p} \left[i  \sin(2 E_p t)  (v^T \gamma^0 w) + (1 -\cos(2 E_p t) ) (v^T \gamma_5\frac{\gamma^j  p^j}{E_p} w)\right]  \delta(\mathbf{p}+\mathbf{q}).        \]
 Here $v$ and $w$  are the constant bispinors present in, respectively,  $ \psi_{\mathbf{p}}$ and $ \psi_{\mathbf{q}}$, see formula (3.5), and   $E_p = \sqrt{m^2 + \mathbf{p}^2}$.  

 The Heisenberg uncertainty relation for the position and the axial momentum has the same form as with the standard momentum \cite{arodz2},
 \[\langle \psi| (\Delta \hat{ x}^j)^2|\psi\rangle \langle \psi |(\Delta \hat{p}_5^k)^2|\psi\rangle \geq  \frac{1}{4} \delta_{jk}, \]  
where  $\Delta \hat{x}^j= \hat{x}^j - \langle\psi| \hat{x}^j|\psi \rangle$, $\;\Delta \hat{p}^k_5 = \hat{p}^k_5 - \langle\psi| \hat{p}^k_5|\psi \rangle$.

\section{General solution of the Dirac equation}
From a mathematical viewpoint, the Dirac equation (3.3), or equivalently Eq.\ (3.4),  is rather simple linear partial differential equation with constant coefficients. It can be solved by the  Fourier transform method.  The standard Fourier transform uses the functions    $\exp(i \mathbf{p} \mathbf{x})$ which are eigenfunctions of the standard momentum   $\hat{\mathbf{p}}$.   In view of the inadequacy of this momentum for the Majorana particle, we prefer an expansion into the eigenfunctions of the axial momentum with the exponential orthogonal matrices $\exp(i \gamma_5 \mathbf{p} \mathbf{x})$.

The eigenfunctions  (3.5)  contain arbitrary real bispinors $v$. At each fixed eigenvalue $\mathbf{p} $ they form real four-dimensional space. We choose as the basis in this space  eigenvectors  of the  real and Hermitian  matrix  $\gamma^0 \gamma^k p^k $, i.e., such $v$ that
\begin{equation} \gamma^0 \gamma^k p^k \:v= E_0 \:v,  \end{equation} 
where the matrices  $\gamma^{\mu}$ have the form  given by (3.1).  
It turns out that the eigenvalues $E_0 = \pm |\mathbf{p}|$.  The eigenvectors have the following form:  for $E_0=|\mathbf{p}|$ 
\begin{equation}
v_1^{(+)}(\mathbf{p}) = \frac{1}{\sqrt{2  |\mathbf{p}|  (  |\mathbf{p}|  - p^2) }} \left(\begin{array}{c} - p^3\\  p^2 - |\mathbf{p}| \\ p^1 \\ 0   \end{array}  \right),    \;\;\; v_2^{(+)}(\mathbf{p}) =  i \gamma_5\: v_1^{(+)}(\mathbf{p}), 
\end{equation}
and for $E_0 = - |\mathbf{p}|$ 
\begin{equation}  
v_1^{(-)}(\mathbf{p}) =  i \gamma^0\: v_1^{(+)}(\mathbf{p}),  \;\;\; v_2^{(-)}(\mathbf{p}) =  i \gamma_5\: v_1^{(-)}(\mathbf{p}) = - \gamma_5 \gamma^0 v_1^{(+)}(\mathbf{p}). 
\end{equation}
These bispinors are real and orthonormal
\[  (v^{(\epsilon)}_j)^T(\mathbf{p}) \: v^{(\epsilon')}_k(\mathbf{p}) = \delta_{\epsilon \epsilon'} \delta_{jk}, \] 
where $\epsilon, \epsilon' = +, -$,  and $j, k = 1,2.$  
  
 Equation (4.1)  is equivalent to \footnote{ $ \hat{E}_0$  should not be confused with the energy operator $\hat{E}= \sqrt{m^2 - \nabla^2}$. We  keep here the notation introduced in \cite{arodz}. }
\[  \hat{E}_0 \:  \psi_{ \mathbf{p}}(\mathbf{x})   =  E_0 \: \psi_{ \mathbf{p}}(\mathbf{x}), \]
where 
\[ \hat{E}_0 = \gamma^0 \gamma^k  \hat{p}^k_5. \]
As shown in \cite{arodz},   $\hat{E}_0$ is related to the standard helicity operator $\hat{\lambda} = S^i\:  \hat{p}^i/ |\hat{\mathbf{p}}|$,  namely
\[ \hat{E}_0 = 2 |\hat{\mathbf{p}}| \hat{\lambda},  \]
where $\: S^j = i \epsilon_{jkl} [\gamma^k, \gamma^l] /8$ are  spin matrices, and $ |\hat{\mathbf{p}}| = \sqrt{\hat{\mathbf{p}}^2} =\sqrt{\hat{\mathbf{p}}_5^2} =  |\hat{\mathbf{p}}_5|$.   Both $\hat{E}_0$ and $\hat{\lambda}$  are observables  (they are real and Hermitian), as opposed to $ S^i$ and $\hat{\mathbf{p}}$ which are not real.  
 Thus, $\hat{E}_0$ is essentially equivalent to the helicity.  The plus sign in (4.2) and the minus in (4.3) correspond to the helicities $+1/2$ and $-1/2$, respectively. 

The expansion of the wave function we start from reads
\begin{equation}\psi(t, \mathbf{x}) = \frac{1}{(2\pi)^{3/2}} \sum_{\alpha=1}^2 \int\!d^3p\: e^{i \gamma_5 \mathbf{p} \mathbf{x}} \left(v_{\alpha}^{( +)}(\mathbf{p})  c_{\alpha}(\mathbf{p},t) +  v_{\alpha}^{( -)}(\mathbf{p})  d_{\alpha}(\mathbf{p},t)\right). \end{equation} 
The time dependence of the axial momentum amplitudes $c_{\alpha}(\mathbf{p},t), \:d_{\alpha}(\mathbf{p},t)$ is determined  by the Dirac equation (3.3).  A series of  mathematical steps  described in \cite{arodz2} leads to the following result
\begin{equation}
\psi(t, \mathbf{x}) = \frac{1}{2(2\pi)^{3/2}} \int\!d^3p\: \left[ \cos(\mathbf{p}\mathbf{x} - E_pt) \: A_+(\mathbf{p})  + \cos(\mathbf{p}\mathbf{x} + E_pt) \:A_-(\mathbf{p})  \right. 
\end{equation} 
\[  \left.  +\sin(\mathbf{p}\mathbf{x} - E_pt) \: B_+(\mathbf{p})  + \sin(\mathbf{p}\mathbf{x} + E_pt) \:B_-(\mathbf{p}) \right], \] 
where 
\[ A_{\pm}(\mathbf{p}) = v^{(+)}_1(\mathbf{p}) A^{1}_{\pm}(\mathbf{p})  +v^{(+)}_2(\mathbf{p}) A^{2}_{\pm}(\mathbf{p}) +v^{(-)}_1(\mathbf{p}) A^{3}_{\pm}(\mathbf{p}) +v^{(-)}_2(\mathbf{p}) A^{4}_{\pm}(\mathbf{p}),   \]  
\[ B_{\pm}(\mathbf{p}) = v^{(+)}_1(\mathbf{p}) B^{1}_{\pm}(\mathbf{p})  +v^{(+)}_2(\mathbf{p}) B^{2}_{\pm}(\mathbf{p}) +v^{(-)}_1(\mathbf{p}) B^{3}_{\pm}(\mathbf{p}) +v^{(-)}_2(\mathbf{p}) B^{4}_{\pm}(\mathbf{p}),   \]  
and 
\[ A^{1}_{\pm} =   (1 \pm \frac{p}{E_p})  c_1(\mathbf{p},0)   \mp   \frac{m}{E_p} d_2(\mathbf{p},0),  \;\;\; A^{2}_{\pm} =   (1 \pm \frac{p}{E_p}) c_2(\mathbf{p},0)  \mp  \frac{m}{E_p} d_1(\mathbf{p},0), \] 
\[  A^{3}_{\pm} =   (1 \mp \frac{p}{E_p}) d_1(\mathbf{p},0)   \pm   \frac{m}{E_p} c_2(\mathbf{p},0), \;\;\;  A^{4}_{\pm} =  (1 \mp \frac{p}{E_p}) d_2(\mathbf{p},0)   \pm   \frac{m}{E_p} c_1(\mathbf{p},0), \]
\[ B^{1}_{\pm} =  -  (1 \pm \frac{p}{E_p})  c_2(\mathbf{p},0)   \mp   \frac{m}{E_p} d_1(\mathbf{p},0),  \;\;\; B^{2}_{\pm} =   (1 \pm \frac{p}{E_p}) c_1(\mathbf{p},0)  \pm  \frac{m}{E_p} d_2(\mathbf{p},0), \] 
\[  B^{3}_{\pm} = -  (1 \mp \frac{p}{E_p}) d_2(\mathbf{p},0)   \pm   \frac{m}{E_p} c_1(\mathbf{p},0), \;\;\;  B^{4}_{\pm} =  (1 \mp \frac{p}{E_p}) d_1(\mathbf{p},0)   \mp   \frac{m}{E_p} c_2(\mathbf{p},0). \]
In these formulas $p \equiv |\mathbf{p}|$,  $E_p = \sqrt{\mathbf{p}^2+ m^2}$, and  $  c_{\alpha}(\mathbf{p},0), d_{\alpha}(\mathbf{p},0)$ are the initial values of the amplitudes given at $t=0$. 
 Let us remind that $\mathbf{p}$   is the eigenvalue of the axial momentum. 

Let us return to the question of energy operator raised in the previous Section.  Because the Hamiltonian $\hat{h}$ is not observable when $m \neq 0$, we have to look for another operator.  Heuristically,  energy in quantum physics   is related to frequency.  This idea can be embodied in the formula  \[\partial_t^2 \psi(t, \mathbf{x})  = -  \hat{E}^2 \psi(t, \mathbf{x}). \]
Inserting here $\psi(t, \mathbf{x})$ given by formula (4.5)  we obtain the condition $\hat{E}^2 = -\nabla^2 + m^2$, from which we would like to determine the energy operator $\hat{E}$.  The simplest  real and Hermitian solution is $\hat{E} = \sqrt{m^2 -\nabla^2}$. The square root can be a multivalued  operation -- in order to avoid misunderstandings let us specify that by $  \sqrt{m^2 -\nabla^2}$  we mean the operator such that 
 \[ \sqrt{m^2 -\nabla^2}\: \psi_{\mathbf{p}}(\mathbf{x}) = \sqrt{m^2 + \mathbf{p}^2} \: \psi_{\mathbf{p}}(\mathbf{x}) \] 
 for all  eigenfunctions  (3.5) of $\hat{\mathbf{p}}_5$. The square root on the r.h.s.  has only non-negative values  by assumption.

Single mode with fixed value $\mathbf{q}$ of the axial  momentum  is obtained by putting in the formulas above \[c_{\alpha}(\mathbf{p},0) = c_{\alpha}\: \delta(\mathbf{p}- \mathbf{q}), \;\;\;\; \:d_{\alpha}(\mathbf{p},0) =  d_{\alpha}\: \delta(\mathbf{p}- \mathbf{q}),\] where $c_{\alpha}, d_{\alpha}$, $\alpha=1,2,$ are constants.  Then,  in the massless case,
\[ A^1_+ = 2c_1, \; A^2_+ = 2c_2, \;  A^3_+= A^4_+=  A^1_- =A^2_- =0, \;A^3_- = 2d_1, \; A^4_- = 2d_2, \] 
\[ B^1_+ =- 2c_2, \; B^2_+ = 2c_1, \;  B^3_+= B^4_+=  B^1_- =B^2_- =0, \;B^3_- =-2d_2, \; B^4_- = 2d_1. \] 
It is clear that the $A_+, B_+$   part on the r.h.s. of formula (4.5)  is independent of the  $A_-, B_-$ part. In particular,   we can put one of them to zero in order  to obtain a plane wave propagating in the direction of $\mathbf{q}$  or $-\mathbf{q}$.  
The massive case is very different --  always  two components propagating in the opposite directions, $\mathbf{q}$ and $-\mathbf{q}$, are present.   If we assume that $A_-=B_-=0$, a simple calculation shows that then also $A_+=B_+=0$, and vice versa. 
Such a pairing of traveling plane waves is one more peculiarity of quantum mechanics of the massive Majorana particle. 

Continuing  the analysis of the single mode,  let  us put  $d_1 =d_2=0$ and keep $c_1$ and $ c_2$ finite.  In the massless case we obtain  plane wave moving in the direction $\mathbf{q}$,  namely
\[ \psi(\mathbf{x}, t) = \frac{1}{(2\pi)^{3/2}}  \left( \cos(\mathbf{q}\mathbf{x} - E_qt) \: (c_1 v^{(+)}_1(\mathbf{q})  +c_2 v^{(+)}_2(\mathbf{q})) \right. \]   \[ \left.  +\sin(\mathbf{q}\mathbf{x} - E_qt) \:(-  c_2 v^{(+)}_1(\mathbf{q})  +c_1 v^{(+)}_2(\mathbf{q}) )   \right). \] 
  In the massive case  all four components in (4.5) do not vanish. 
However,   the amplitudes   $A_-$ and $ B_-  $ of the $- \mathbf{q}$  components are negligibly small in the high energy limit  ($m/E_q \ll 1$). In this limit  
\[ A^1_+ \approx 2 c_1, \; A^2_+ \approx 2 c_2, \;  A^3_+ =\frac{m}{E_q} c_2 , \; A^4_+ = \frac{m}{E_q} c_1,  \]  \[ B^1_+ \approx -2 c_2, \; B^2_+ \approx 2 c_1, \;  B^3_+ =\frac{m}{E_q} c_1 , \; B^4_+ = - \frac{m}{E_q} c_2,  \]
and \[  A^1_- \approx \frac{m^2}{2 E_q^2} c_1, \;A^2_- \approx \frac{m^2}{2E_q^2} c_2, \;A^3_- = -\frac{m}{E_q}c_2, \; A^4_- = \frac{m}{E_q}c_1, \] 
\[  B^1_- \approx -\frac{m^2}{2 E_q^2} c_2, \;B^2_- \approx \frac{m^2}{2E_q^2} c_1, \;B^3_- = -\frac{m}{E_q}c_1, \; B^4_- = \frac{m}{E_q}c_2. \] 
On the other hand, in the limit of small energies ($E_q \approx m$) magnitudes of the  $\mathbf{q}$ and $-\mathbf{q}$ components are approximately equal, 
\[A^1_{\pm} \approx c_1, \; A^2_{\pm} \approx  c_2, \;  A^3_{\pm} \approx \pm c_2 , \; A^4_{\pm} \approx \pm  c_1 ,   \]   and  \[ B^1_{\pm} \approx -c_2, \; B^2_{\pm} \approx  c_1, \;  B^3_{\pm} \approx  \pm c_1, \; B^4_{\pm} \approx  \mp c_2.  \]

\section{ The relativistic invariance }
Relativistic transformations of the Majorana bispinor have the form (2.7),  where now 
$S(L) = \exp(\omega_{\mu\nu} [\gamma^{\mu}, \gamma^{\nu}]/8)$, where the matrices $\gamma^{\mu}$ have the form given by (3.1).  Our goal is to check which irreducible representations of the Poincar\'e group are hidden in the space of real solutions of the Dirac equation (3.3), if any.  

Instead of  $\psi(t, \mathbf{x})$ we shall consider its  counterpart in the axial momentum representation  -- the real bispinor  $v(\mathbf{p},t)$ introduced as follows
 \begin{equation}
\psi(t, \mathbf{x}) =\frac{1}{ (2\pi)^{3/2}}  \int\! \frac{d^3p}{E_p}\: e^{i \gamma_5 \mathbf{p} \mathbf{x}} \:   v(\mathbf{p}, t), 
 \end{equation}
where    $E_p = \sqrt{m^2 + \mathbf{p}^2}$.   Equation (3.4)  gives time evolution equation for $v$  
\begin{equation}
\dot{v}(\mathbf{p}, t)  = - i \gamma^0 \gamma^k \gamma_5 p^k v(\mathbf{p}, t) - im \gamma^0 v(-\mathbf{p}, t).  
\end{equation}
We have $v(-\mathbf{p}, t)$   in the last term on the r.h.s.  because 
$\gamma^0 \exp(i \gamma_5 \mathbf{p} \mathbf{x}) =   \exp(- i \gamma_5 \mathbf{p} \mathbf{x}) \gamma^0.$
 From Eq.\  (5.2)   we obtain equation
\[ \ddot{v}(\mathbf{p}, t) = - E_p^2 v(\mathbf{p}, t),  \]  which has general solution in the form 
\begin{equation}
v(\mathbf{p}, t) = \exp(-i\gamma_5 E_p t) v_+(\mathbf{p})  +   \exp(i\gamma_5 E_p t) v_-(-\mathbf{p}), 
\end{equation}
where $v_{\pm}$ are arbitrary real bispinors (we write   $v_-(-\mathbf{p})$  for later convenience). 
Formulas (5.3) and  (5.1) give 
\begin{equation}
\psi(\mathbf{x}, t) =\frac{1}{ (2\pi)^{3/2}}  \int\! \frac{d^3p}{E_p}\:\left( e^{i \gamma_5 (\mathbf{p} \mathbf{x}-E_pt)}\:   v_+(\mathbf{p})  +   e^{-i \gamma_5 (\mathbf{p} \mathbf{x}-E_pt)}\:   v_-(\mathbf{p})   \right).
 \end{equation}
 We have changed   the integration variable to $- \mathbf{p}$  in the $v_-$ term. 
 Furthermore,  Eq.\ (5.2)  implies the  following relations
\begin{equation}
E_p  \gamma_5 v_{\pm}(\mathbf{p}) = \gamma^0 \gamma^k p^k \gamma_5 v_{\pm}(\mathbf{p}) \pm m \gamma^0  v_{\mp}(\mathbf{p}) . 
\end{equation}

Transformation law (2.7) with $a=0$  applied to  solution  (5.4) gives  Lorentz transformation of the bispinors $ v_{\pm}(\mathbf{p})$,    
\begin{equation}
v_{\pm}^{'}(p) =   S(L) \:v_{\pm}(L^{-1}p).
\end{equation}
Here we use the four-vector $p$ instead of $\mathbf{p}$  in order to simplify notation:   $v_+(p) \equiv v_+(\mathbf{p})$,  where $p^0 =E_p$.  
In the case of space-time  translations $x'  = x + a  $ we obtain  
\begin{equation}  v'_{\pm}(\mathbf{p}) = e^{\pm i \gamma_5  p a}   v_{\pm}(\mathbf{p}). \end{equation}

Further steps depend on whether the particle is massive or massless.  The presented below discussion of the massive case is based on  Section 4 of \cite{arodz2}, where all omitted details can be found.  The massless case is not covered in that paper -- it is presented below for the first time.

\subsection{The  massive  Majorana particle }
 In this  case  $v_-(\mathbf{p}) $  can be expressed by $v_+(\mathbf{p})$, see (5.5).  Using formula (5.4)  we find that the scalar product $\langle\psi_1|\psi_2\rangle = \int \!d^3x\:\psi_1^T(t, \mathbf{x}) \psi_2(t, \mathbf{x}) $  is equal to  
\begin{equation} \langle \psi_1 | \psi_2 \rangle = \frac{2}{m^2}  \int\!\frac{d^3p}{E_p}\: \overline{v_{1+}(\mathbf{p})}\: ( \gamma^0 E_p - \gamma^k p^k)\: v_{2+}(\mathbf{p}),\end{equation} 
where $\overline{v_{1+}(\mathbf{p})} = v_{1+}^T(\mathbf{p}) \gamma^0$,  and $v_{1+}$ $(v_{2+}) $ corresponds to $\psi_1$ $(\psi_2)$ by formula (5.4).  The form (5.8) of the scalar product is  explicitly Poincar\'e  invariant and time independent.  
 
Transformations  (5.6), (5.7)  are unitary with respect to scalar product  (5.8). Thus, we have found certain real unitary, i.e., orthogonal,  representation of the Poincar\'e  group.  In order to determine the spin quantum number for this representation, we 
recast it to  the standard  form with the Wigner rotations  \cite{raczka}.
First,  we choose the standard momentum $\stackrel{(0)}{p} = (m, 0,0,0)^T$, where $m >0$,  and a Lorentz boost $H(p)$,   $H(p) \stackrel{(0)}{p} = p$.  At each  $p$ we introduce the basis  of real bispinors,
\begin{equation} v_i(p) = S(H(p)) v_i(\stackrel{(0)}{p}),\end{equation} 
  where $i = 1, 2, 3, 4$.   Here $ v_i(\stackrel{(0)}{p})$ is a basis at $ \stackrel{(0)}{p}$ such that  $ v^T_i(\stackrel{(0)}{p})  v_k(\stackrel{(0)}{p}) = \delta_{ik}/m$.   Actually,  we assume that  this basis has the  Kronecker form, i.e.,   the $i$-th component of the bispinor  $ v_k( \stackrel{(0)}{p} )$  is equal to $\delta_{ik}/\sqrt{m}$.   The factor $m$ is present for dimensional reason.
In (5.9) the four-momentum    notation is used,  as in (5.6).  We write  $v_+(p)$  
in this basis,  \[v_+(p) = a^i(p) v_i(p).\]   The amplitudes $a^i(p)$, $i=1, 2, 3, 4$,  are real and dimensionless.  The scalar product (5.8)  is equal to \begin{equation}  \langle \psi_1| \psi_2 \rangle = \frac{2}{m^2} \int \! \frac{d^3 p}{E_p} \: a_1^i(p) a_2^i(p), \end{equation}
where    $a^i_1, a^i_2$  correspond to $\psi_1, \psi_2$, respectively. 
The remaining steps  are rather technical.    For detailed  description of them  we refer the reader to the paper \cite{arodz2}.  Below we  cite the main results.

It turns out that  Lorentz transformations  (5.6)  imply the following transformation of the  amplitudes  $a^i$
\begin{equation}
a^{'k}(p) =  S_{ki}({\cal R}(L,p))  a^i(L^{-1}p),
\end{equation}
where  ${\cal R}(L,p) =H^{-1}(p) L H(L^{-1}p)$ is the Wigner rotation,  and  $S_{ki}  $ are  the matrix elements of the matrix $S(L)$  introduced in formula (2.7). 
In the case of translations 
\begin{equation}
a^{'k}(p) =  ( e^{i \gamma_5 pa})_{ki} a^i(p).   
\end{equation}

For an  arbitrary rotation $R$, including the Wigner rotation,   the matrix $S(R)$ has the form 
\[  S(R) =  \exp(\frac{1}{2} (\omega_{12} \gamma^1\gamma^2 + \omega_{31} \gamma^3\gamma^1 +  \omega_{23} \gamma^2\gamma^3)).  \]
It can be shown that there exist a real orthogonal matrix ${\cal O}$  such that 
\begin{equation}  {\cal O} S(R) {\cal O}^{-1} =  \hat{T},
\end{equation}
where the four by four real matrix $ \hat{T}$  has the form 
\begin{equation} \hat{T} = \left( \begin{array}{cccc} \alpha' & - \alpha'' & -\beta' &  \beta'' \\  \alpha'' & \alpha' & -\beta'' &  -\beta' \\  \beta' &  \beta'' & \alpha' & \alpha'' \\  -\beta''&  \beta' & -\alpha'' & \alpha' 
\end{array}  \right).  \end{equation} 
The  parameters $\alpha', \alpha'', \beta',  \beta''$  are certain  functions of  $\omega_{12}$, $\omega_{31}$, $\omega_{23}$.  

In the last step, we recognize in the matrix   $\hat{T}$  the real form of  the spin 1/2  representation  $T(u)$ of $SU(2)$ group.  This  representation is  given by the transformations  $T(u) \xi = u \xi$,  where $u\in SU(2)$ and $\xi$ is  a two-component spinor, in general complex.   Its real form is obtained  simply by using  the real and imaginary parts.  Let us take  
\[  u = \left( \begin{array}{cc} \alpha & \ -\beta \\ \beta^* & \alpha^*   \end{array}   \right), \;\;\; \xi = \left( \begin{array}{c} \xi_1 \\   \xi_2 \end{array}  \right), \] 
where $\alpha= \alpha' + i \alpha'',  \; \beta = \beta' + i \beta'', \;  \xi_1= \xi_1' + i \xi_1'', \; \xi_2= \xi_2' + i \xi_2'', $ and 
$ \alpha \alpha^* + \beta \beta^* =(\alpha')^2 +(\alpha'')^2 + (\beta')^2 +(\beta'')^2 =1.$
The real forms of $\xi$ and $u$  read   \[
\vec{\xi} = \left(  \begin{array}{c} \xi_1' \\ \xi_1'' \\ \xi_2' \\ \xi_2''  \end{array}   \right), \;\;\; \hat{T}(u) = \left( \begin{array}{cccc} \alpha' & - \alpha'' & -\beta' &  \beta'' \\  \alpha'' & \alpha' & -\beta'' &  -\beta' \\  \beta' &  \beta'' & \alpha' & \alpha'' \\  -\beta''&  \beta' & -\alpha'' & \alpha' 
\end{array}  \right).  \]
The real form of the spinor $u \xi$ is equal to $ \hat{T}(u) \vec{\xi}.$  

We conclude that  the representation  (5.11) is equivalent to the real form of the spin 1/2 representation $T(u)$ of  $SU(2)$ group.  Thus, the unveiled representation of the Poincar\'e group  is the spin 1/2, $m>0$, representation. 
Let us emphasize that we have obtained  just one such  representation. For comparison,  in the  case of  Dirac particle two spin 1/2 representations  are present.   The representations usually reappear in quantum field theory.  Single representation in  the Majorana  case  would  correspond  to a single spin 1/2 particle. In the Dirac case there are two representations because there is particle and its anti-particle.

\subsection{ The   massless Majorana particle}
We again use  formula (5.4) and transformations (5.6), (5.7).  The difference with the massive case is that now  the bispinors $v_+$, $v_-$ are independent. Relations (5.5)   with $m=0$  become constraints for them, namely 
\begin{equation}
(\gamma^0 E_p - \gamma^k p^k)  v_{\pm}(\mathbf{p})=0, 
\end{equation}
where $E_p = |\mathbf{p}|$.
Linear conditions (5.15) define two subspaces of real bispinors   $v_+$, $v_ -$ which are two-dimensional.  Each subspace spans the same representation (5.6),  (5.7). 
It turns out that these representations are irreducible, orthogonal, and  characterized by the helicities $\pm 1/2$.  The reason for the opposite signs of the helicities in spite of the same transformation law is that the axial momenta corresponding to  $v_+(\mathbf{p})$ and  $v_-(\mathbf{p})$ are $\mathbf{p}$ and $- \mathbf{p}$, respectively,  because of the opposite signs in the two exponents in formula (5.4). 

One can easily show that   general solution of conditions (5.15)   has the form
\begin{equation} v_{\pm}(\mathbf{p}) = i ( \gamma^0 |\mathbf{p}| - \gamma^k p^k)\:  w_{\pm}(\mathbf{p}),
 \end{equation}  
where  real bispinors $ w_{\pm}(\mathbf{p}))$  are arbitrary.  The crucial fact here is nilpotency of the matrix  on the l.h.s. of  conditions (5.15),
\[    ( \gamma^0 |\mathbf{p}| - \gamma^k p^k)^2=0. \]
For a given  $v_{\pm}(\mathbf{p})$   formula (5.16) determines  $w_{\pm}(\mathbf{p})$ up to a gauge transformation of the form
\begin{equation} w_{\pm}^{’}(\mathbf{p})  =  w_{\pm}(\mathbf{p})  +  i ( \gamma^0 |\mathbf{p}| - \gamma^k p^k) \:\chi _{\pm}(\mathbf{p})  \end{equation}
with arbitrary real  bispinors $\chi_{\pm}(\mathbf{p})$. 

Inserting (5.16) in formula (5.4) we obtain the following  formula   for the scalar product (3.2)
\[ \langle \psi_1 | \psi_2 \rangle = 2 \int\! \frac{ d^3p}{|\mathbf{p}|} \: \left[  \overline{w_{1+}(\mathbf{p})}\: ( \gamma^0 |\mathbf{p}| - \gamma^k p^k)\: w_{2+}(\mathbf{p}) \right. \;\;\;\; \;\;\;\;\;\;\;\;\;\;\] \begin{equation}    \hspace*{2cm}\;\;\;\;\;\;\;\;\;\; \;\;\;\;\;\;\;\;\;\; \;\;\;\;\;\;\;\;\;\;\left. + \:  \overline{w_{1-}(\mathbf{p})}\: ( \gamma^0 |\mathbf{p}| - \gamma^k p^k)\: w_{2-}(\mathbf{p})\right].  \end{equation}
Notice that  the scalar product is invariant with respect to  gauge  transformations (5.17).  

There is a caveat concerning the r.h.s. of formula (5.18). Namely, it should not be considered as  scalar  product of  $w$'s, but rather as scalar product of equivalence classes of which the concrete $w$'s are mere representatives. The equivalence class contains all  bispinors $w_{+}(\mathbf{p})$  (or $ w_{-}(\mathbf{p})$) related to each other by gauge transformations  (5.17). All they give the same   $v_{\pm}(\mathbf{p})$ and  $\psi(t, \mathbf{x})$.   The r.h.s.  of formula  (5.18) does not fulfill the requirement that  for $w_{2\pm}(\mathbf{p}) = w_{1\pm}(\mathbf{p})$   it  vanishes  if and only if $w_{1\pm}(\mathbf{p}) =0$  -- the property of any true scalar product.    The r.h.s. of formula (5.18)  vanishes for any $w_{\pm}$ of the form $ w_{\pm}(\mathbf{p}) =  i ( \gamma^0 |\mathbf{p}| - \gamma^k p^k) \:\chi_{\pm}(\mathbf{p})$. All such  $w_{\pm}$  are gauge equivalent to $w_{\pm}=0$ and they give $\psi(t, \mathbf{x})=0$. 

We assume that Lorentz transformation of $w_{\pm}$ have the following form
\begin{equation}  
w'_{\pm}(p) =   S(L) \:w_{\pm}(L^{-1}p). 
\end{equation}
It implies  transformation law  (5.6)  for  $v_{\pm}$ given by formula (5.16). 
   In the case of translations
 \begin{equation} 
 w'_{\pm}(p) =  e^{\mp i \gamma_5  p a}   w_{\pm}(\mathbf{p}).
 \end{equation}
  Notice that we could allow for  certain gauge transformations on the r.h.s.'s of these  formulas. 
  
   Scalar product (5.18)  is invariant with respect to transformations  (5.19), (5.20).  Therefore, we have two independent unitary (i.e., orthogonal) representations   of the Poincar\'e group.  In order to  identify these representations, we check the related  representations of  the so called little group \cite{raczka}.  In the massless case  the standard momentum
is  $\stackrel{(0)}{p} = (\kappa, 0,0,\kappa)^T,$ where $\kappa>0$ is fixed. The pertinent little group, called $E(2)$,  is the maximal subgroup of the Lorentz group which leaves the standard momentum invariant. It is  three dimensional, and it includes spatial rotations around  $\stackrel{(0)}{\mathbf{p}} = (0,0, \kappa)^T$ as well as  certain combinations of Lorentz boosts and rotations  \footnote{  In the case of massive Majorana particle the little group is the $SO(3)$ subgroup of the Lorentz group.}.  Unitary irreducible representations of $E(2)$ are either infinite dimensional or one-dimensional (over complex numbers) \cite{raczka}. 

In the considerations presented below we concentrate on  the bispinors $w_+$.  Parallel considerations for $w_-$ are essentially identical.  Let us introduce a basis   $w_i(\stackrel{(0)}{p}),$ $ i=1,  2, 3, 4,$  of bispinors at $\stackrel{(0)}{p}$.   Applying Lorentz boosts  $H_0(p)$, which transform $ \stackrel{(0)}{p}$  into $p$, $H_0(p) \stackrel{(0)}{p} =p$, where $(p^0)^2 - \mathbf{p}^2=0$ and $ p^0 >0$, we obtain a basis $w_i(p)$ at each  $p$  belonging to the upper light-cone, 
\begin{equation}
w_i(p) =  S(H_0(p)) w_i( \stackrel{(0)}{p}).
\end{equation}
We decompose $w_+(p)$ in this basis,
\[ w_+(p) = w_i(p) c^i(p).  \] 
Lorentz transformations  (5.19) of bispinors are equivalent to certain transformations of the amplitudes  $c^i(p)$ which give a representation of the little group $E(2)$. The form of these transformations is deduced  from (5.19) as follows. First, 
\[ w'_+(p) = w_k(p) c^{'k}(p) = S(L) w_+(L^{-1}p) =   c^i(L^{-1}p)\: S(L) \: w_i(L^{-1}p) \]
\[ = c^i(L^{-1}p) \:S(H_0(p)) \: S(H_0^{-1}(p) L H_0(L^{-1}p)) \:w_i( \stackrel{(0)}{p}). \] 
Next, we notice that the Lorentz transformation  $H_0^{-1}(p) L H_0(L^{-1}p)$  --   let us denote it by ${\cal E}(L,p)$  -- leaves the standard momentum $ \stackrel{(0)}{p}$ invariant, hence it belongs to the little group $E(2)$.  We decompose  bispinor 
$S({\cal E}(L,p)) w_i( \stackrel{(0)}{p})$ in the basis $w_k(\stackrel{(0)}{p})$, 
\begin{equation} S({\cal E}(L,p)) w_i( \stackrel{(0)}{p}) =  D_{ki}({\cal E}(L,p)) w_k(\stackrel{(0)}{p}), \end{equation}
and write
\[  w_k(p) c^{'k}(p) = c^i(L^{-1}p)  D_{ki}({\cal E}(L,p)) w_k(p). \]
We see from this formula  that 
\begin{equation}
  c^{'k}(p) =  D_{ki}({\cal E}(L,p))\: c^i(L^{-1}p).
  \end{equation}
It is clear that the  Lorentz transformation (2.7) of the Majorana bispinors $\psi(x)$ follows from  transformation (5.23) of the amplitudes $c^i(p)$ (in the massless case,  of course). 

Let us  consider transformation (5.23)  when $p= \stackrel{(0)}{p}$, and   $L= R(\theta)$ is a  rotation around the vector  $\stackrel{(0)}{\mathbf{p}} = (0, 0, \kappa)^T$.   Such $L$    belongs to the $E(2)$  group. In this case (5.22) and (5.23) read  
\begin{equation}
S(R(\theta)) w_i( \stackrel{(0)}{p}) =  D_{ki}(R(\theta)) w_k(\stackrel{(0)}{p}),
\end{equation}
\begin{equation}
   c^{'k}(\stackrel{(0)}{p}) =  D_{ki}(R(\theta))\: c^i(\stackrel{(0)}{p}).
  \end{equation}
 For the rotations around the third axis,  \[S(R(\theta)) =\exp(\gamma^1\gamma^2 \theta/2)  = \cos(\theta/2) I + \gamma^1 \gamma^2 \sin(\theta/2), \] where $\theta$ is the angle, and $\gamma^1, \gamma^2$ are given by (3.1).  We obtain 
 \begin{equation}
 S(R(\theta)) = \left( \begin{array}{cccc}  \cos(\theta/2) & 0 & 0& \sin(\theta/2) \\ 0 &\cos(\theta/2) & \sin(\theta/2) & 0  \\  0 &-  \sin(\theta/2) &\cos(\theta/2) &0 \\-  \sin(\theta/2)  &0 & 0& \cos(\theta/2)          \end{array} \right).
\end{equation}

It remains to specify the basis $w_k(\stackrel{(0)}{p})$. When doing this  we should take into account the fact that not all directions in the bispinor space are relevant for physics  because of gauge transformations (5.17).  In  the case at hand,  they have the following form
\begin{equation}
w'_+(\stackrel{(0)}{p}) = w_+(\stackrel{(0)}{p}) +  \kappa (\chi^4 - \chi^3) e_3 + (\chi^2 - \chi^1) e_4,
\end{equation}
where $\chi^i$ are arbitrary real numbers (components of the bispinor $\chi_+(\stackrel{(0)}{p})$  present in (5.17)), and 
$e_3 = (1,1,0,0)^T, \: e_4= (0,0,1,1)^T$. It is clear that $e_3$ and $ e_4$ give the nonphysical directions in the space of bispinors $w_+$ at $\stackrel{(0)}{p}$.  The remaining two directions -- the physical ones -- are given by $e_1= (1, -1, 0,0)^T$ and $e_2 = (0,0,1,-1)^T.$  Thus, there is a natural choice for the basis $w_k(\stackrel{(0)}{p})$, namely \[ w_k(\stackrel{(0)}{p}) = e_k. \] 
Only the amplitudes $c^1(\stackrel{(0)}{p}), \: c^2(\stackrel{(0)}{p})$ are physically interesting.

Using (5.26) and (5.24) we easily compute  $D_{ki}(R(\theta))$ for $i, k =1,2.$ The transformations (5.25)  have now the form
\[ c^{'1}(\stackrel{(0)}{p}) = \cos(\theta/2) c^1(\stackrel{(0)}{p}) - \sin(\theta/2) c^2(\stackrel{(0)}{p}), \]  \[   
 c^{'2}(\stackrel{(0)}{p}) = \sin(\theta/2) c^1(\stackrel{(0)}{p}) + \cos(\theta/2) c^2(\stackrel{(0)}{p}).  \]
These transformations are the real version of  the following transformations of the complex  amplitude $z(\stackrel{(0)}{p}) = c^1(\stackrel{(0)}{p}) +i c^2(\stackrel{(0)}{p})$: 
\[ z'(\stackrel{(0)}{p}) = \exp(i \theta/2) z(\stackrel{(0)}{p}). \]  Such transformations are characteristic for a massless particle with helicity 1/2.   

 Calculations  for the $w_-$ bispinors  give the same formulas for  transformations, but the helicity is equal to -1/2, because the standard vector $\stackrel{(0)}{\mathbf{p}} = (0, 0, \kappa)^T$ corresponds to the axial momentum 
$-\!\!\!\stackrel{(0)}{\mathbf{p}}$,  recall $v_-(-\mathbf{p})$ in formula (5.3).

\section{Remarks}
{\bf1.} We have outlined the basic structure of  the relativistic  quantum mechanics of the Majorana particle. List of its specific elements includes: the Hilbert space over $\mathbb{R}$,  not over $\mathbb{C}$ as for other particles;  the lack of the standard momentum operator and  the appearance of the axial momentum  with its peculiarities present in the case of massive particle;  multiplicities of real orthogonal irreducible representations of the Poincar\'e  group consistent with the expected lack of  anti-particle in quantum theory of the Majorana field. One more item,  not discussed  here, is a relation with quaternions \cite{arodz2}.   It is clear that it is very interesting  theory, worth further  studies.

{\bf 2.} Formula  (4.5) can be used for studying  time evolution of wave packets with specified initial content of the axial momentum \cite{arodz3}. This  is  rather interesting problem  because the axial momentum is not constant in time when $m\neq 0$, hence we do not have  any simple intuitions about the time evolution. Another topic one would like to know more about is  behavior of the Majorana particle in external potentials, which also can be studied with  use of the axial momentum.

{\bf 3.} The results of the analysis of relativistic invariance illustrate the well-known fact that  the theory of  massless  particle is not a simple $m \rightarrow 0$ limit of  the theory of massive particle. In particular,  in the massive case there is a single orthogonal irreducible representation of the Poincar\'e group, while for the massless particle we have two  representations.  Moreover,   the case of  massless Majorana particle is distinguished by the presence of the  gauge structure, as shown in Section 5.2.   Gauge structures behind massless photons and gluons are well-known, but its presence also in the case of Majorana particle is a surprise.

\section{Acknowledgement} The author acknowledges  a support from the Marian Smoluchowski Institute of Physics, Jagiellonian University,  Contract No.  337.1104.112.2019.

\end{document}